# Twisted photon entanglement through turbulent air across Vienna

Mario Krenn[1,2,*], Johannes Handsteiner[1,2], Matthias Fink[1,2], Robert Fickler[1,2,3], Anton Zeilinger[1,2,*]

[1]Vienna Center for Quantum Science and Technology (VCQ), Faculty of Physics, University of Vienna, Boltzmanngasse 5, A-1090 Vienna, Austria.
[2]Institute for Quantum Optics and Quantum Information (IQOQI), Austrian Academy of Sciences, Boltzmanngasse 3, A-1090 Vienna, Austria.
[3]present address: Department of Physics and Max Planck Centre for Extreme and Quantum Photonics, University of Ottawa,Ottawa, K1N 6N5, Canada
[*]correspondence to mario.krenn@univie.ac.at and anton.zeilinger@univie.ac.at

**Photons with a twisted phase front can carry a discrete, in principle unbounded amount of orbital angular momentum (OAM). The large state space allows for complex types of entanglement, interesting both for quantum communication and for fundamental tests of quantum theory. However, the distribution of such entangled states over large distances was thought to be infeasible due to influence of atmospheric turbulence, indicating a serious limitation on their usefulness. Here we show that it is possible to distribute quantum entanglement encoded in OAM over a turbulent intra-city link of 3 kilometers. We confirm quantum entanglement of the first two higher-order levels (with OAM=$\pm 1\hbar$ and $\pm 2\hbar$). They correspond to four new quantum channels orthogonal to all that have been used in long-distance quantum experiments so far. Therefore a promising application would be quantum communication with a large alphabet. We also demonstrate that our link allows access to up to 11 quantum channels of OAM. The restrictive factors towards higher numbers are technical limitations that can be circumvented with readily available technologies.**

Long-distance quantum entanglement with photons opens up the possibility to test fundamental properties of quantum physics in regimes not accessible in lab scale experiments, it can be used for quantum communication between widely separated parties, and it is the basis of quantum repeaters as nodes in a global quantum network. As the polarization of photons is easily controllable and resistant against atmospheric turbulences, it has been successfully used in a variety of different long-distance quantum experiments [1-4]. However, polarization of photons resides in a two-dimensional state-space, restricting the complexity of entangled states both for certain quantum communication tasks and for fundamental tests.

In contrast to polarization, the orbital-angular momentum (OAM) modes of photons have an unbounded state-space. Photons carrying OAM have a twisted wave front with a phase that varies from 0 to $2\pi\ell$ in the azimuthal direction. Here, $\ell$ is an integer which stands for the topological charge, and $\ell \cdot \hbar$ is the OAM of the photon. Such states can carry larger amount of information per photon. It also allows more complex types of non-classical correlations, such as entanglement of large quantum numbers [5] or high-dimensional entanglement [6-10]. However, the possibility of more complex entangled quantum states poses a substantial challenge due to the negative influence of atmospheric turbulences on such modes. Several theoretical [11-17] and lab-scale [18-20] studies investigated the effect of turbulence on entanglement encoded in the OAM of photons, and many others explore the influence of turbulence on OAM modes in general [21-24]. Only one quantum experiment was carried out beyond the lab-scale, by performing a quantum communication protocol over 210 meters using a polarization-OAM hybrid system [25]. It was located in a large hall to minimize the disturbing effects of turbulence. So far, no experiment at the quantum level has been performed in a long-distance turbulent real-world environment, and quantum entanglement has not yet been demonstrated beyond the lab-scale with photons carrying OAM.

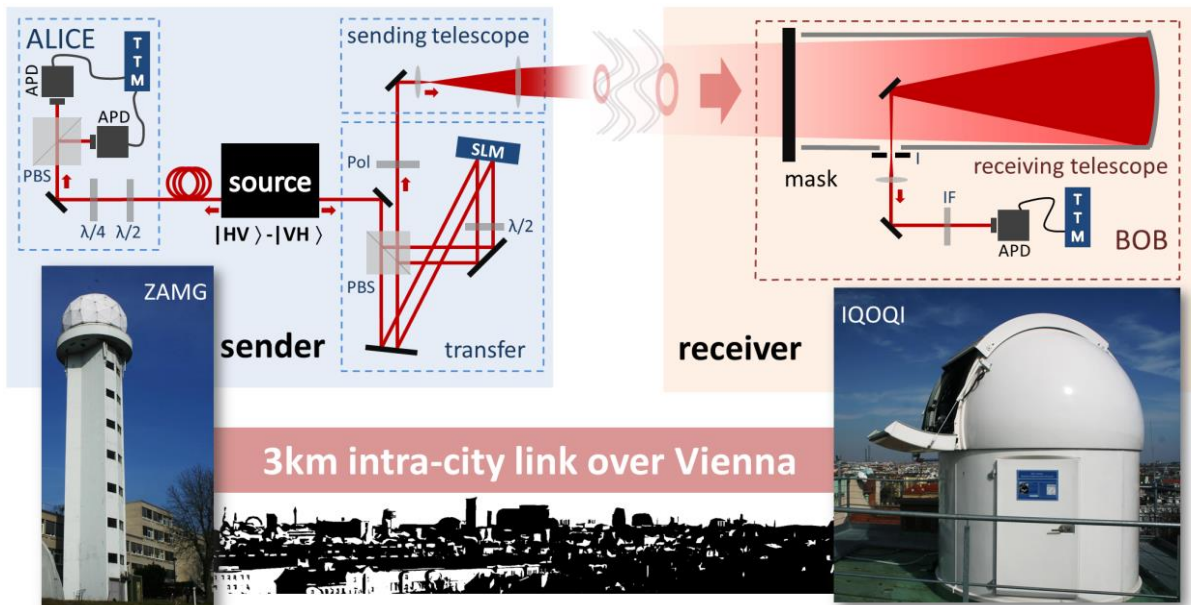

**Figure 1:** Sketch of the experimental setup. The experiment takes place at two locations separated by 3 kilometers. The sender is located in a radar tower of ZAMG (*Zentralanstalt für Metrologie und Geodynamik*), the receiver is at the rooftop of our institute IQOQI (*Institute for Quantum Optics and Quantum Information*). <u>Left</u>: At the sender, we have a high-fidelity Sagnac-type polarization entanglement source. One photon is transferred to the OAM degree of freedom, using an interferometric scheme [5, 27]: In it, the photon's path is separated according to its polarization at a polarizing beam splitter (PBS) and transformed to an OAM value depending on its path using a spatial light modulator (SLM, *Hamamatsu LCOS-SLM*). After recombination of the paths, the transfer is completed by deleting the polarization information with a polarizer (Pol). Subsequently, the photon wave front is expanded and sent to the transceiver with a high-quality lens. Meanwhile the second photon of the entangled pair is delayed in a 30 meter fiber to ensure the transfer and sending of the first photon before the second photon is detected. After the fiber the second photon is measured using a half-wave plate ($\lambda/2$) or a quarter-wave plate ($\lambda/4$) – depending on the basis in question –, a PBS and two avalanche photon detectors (APDs). The detection times of the photons are recorded with a time tagging module (TTM). <u>Right</u>: At the receiver, the transmitted photons are collected by a Newton-type telescope with a primary mirror of 37 cm diameter. In front of the primary mirror, opaque masks with symmetric slit patterns are used to perform mode measurements (see Fig.2). An iris (I) and a 3nm band pass filter (IF) were used to minimize background light. The photons are detected with an APD, and time tagged with a TTM. Coincidences are then extracted by comparing the time tagging information from both locations.

Recently, an investigation of the transmission of classical light with different spatial modes across an intra-city link indicated that the phase of OAM superpositions is well conserved during the transmission, hinting that the distribution of quantum entanglement encoded in OAM might be possible [26]. Here, we present the results of an experiment, in which we show that entanglement distribution with spatial modes is indeed possible over a turbulent intra-city link.

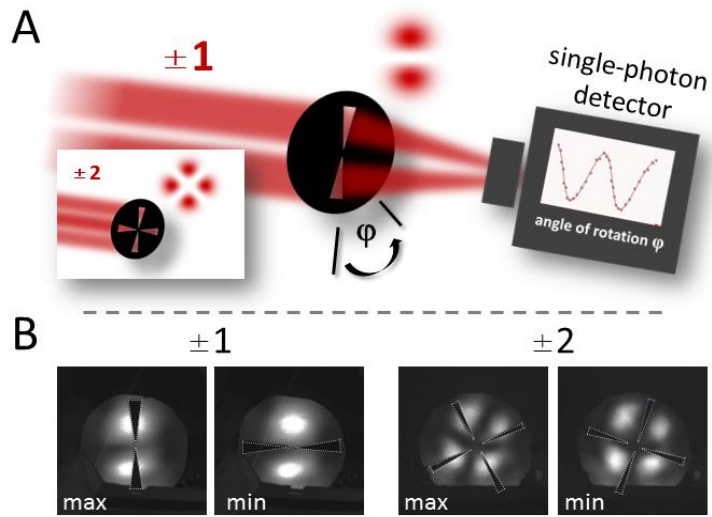

**Figure 2**: Principle of the measurement technique. The superposition of two LG modes with opposite $\ell$ has $2\ell$ minima and maxima in a ring. The angular orientation depends on the relative phase. We use a mask, which resembles the symmetry of the beam, to measure correlations. <u>A</u>: An incoming beam hits the mask. For $\ell=\pm1$, an opaque mask with two transparent slits is used to measure different superposition states. A detector after the mask collects all photons passing the slits. The superposition of $\ell=\pm2$ has four paddles, thus we use a mask with four slits. <u>B</u>: Images of an alignment laser beam at the mask (mounted at the telescope at IQOQI, slits are highlighted) after 3 kilometer transmission. The laser is in a superposition of $\ell=\pm1$ and $\pm2$. The angular position of the mask is set to the maximum and to the minimum. In the entanglement experiment, we see the fringes only in coincidences.

The experimental setup can be divided into four main parts (Fig. 1): the source of polarization entanglement, the transfer of one photon from polarization to the OAM degree of freedom, Alice's polarization analysis and Bob's OAM measurement after transmission. The sender (Alice) and the receiver (Bob) are at different physical locations 3 kilometers apart. The sender is located in a ~35 meter high radar tower of *Zentralanstalt für Metrologie und Geodynamik* (ZAMG – *Central Institute for metrology and geodynamics*). There we use a high-fidelity high-brightness polarization entanglement source [28, 29] with an uncorrected average visibility of ~97.5%. Photon A is unchanged, whereas photon B's polarization state is transferred interferometrically to an OAM state [5]. After the transfer the generated hybrid-entangled quantum state can be written as

$$|\psi\rangle = \tfrac{1}{\sqrt{2}}\bigl(|H\rangle_A|+\ell\rangle_B + |V\rangle_A|-\ell\rangle_B\bigr). \qquad (1)$$

To ensure that photon B is transferred and sent before photon A is measured, we delay photon A by a 30 meter fiber. It ensures that the detection of the two entangled photons is space-like separated: Any signal from one measurement to the other would need to propagate faster than the speed of light to influence the result. Afterwards the polarization of photon A is measured with a two-output polarization analyzing station. Each detection event is time-stamped and recorded with a time tagging module (*Austrian Institute Of Technology TTM 8000*). The OAM-encoded photon B is magnified to a Gaussian beam waist of 11 millimeters and send through turbulent atmosphere with a high-quality lens using a focal length of f=30cm to the receiver 3 kilometers away.

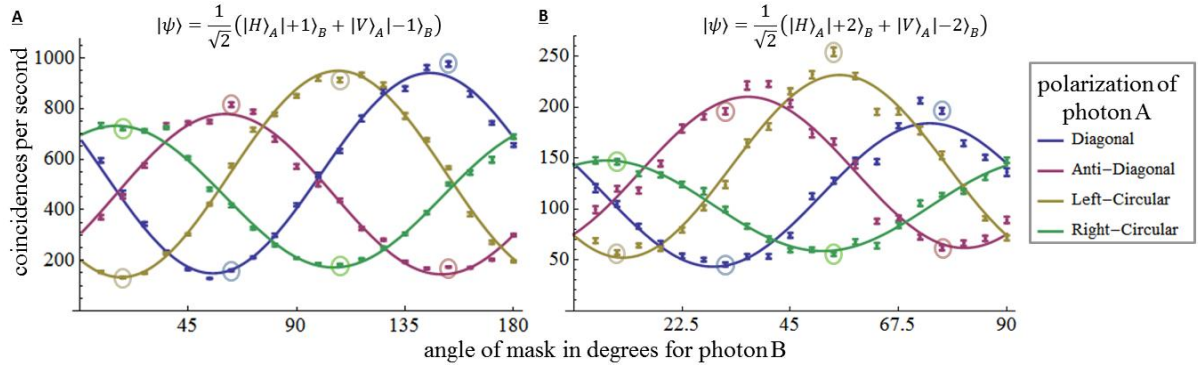

**Figure 3:** Coincidences between the transmitted photon encoded in OAM and the locally measured polarization photon. For four different polarization settings at Alice's photon A, coincidences were recorded for 20 different angular positions of the mask at Bob's receiver. Error bars are the standard deviation of the mean, calculated without assumptions about the underlying photon distribution from raw data by splitting the whole measurement time into 1-second intervals. The circles indicate the data used to calculate the entanglement witness. The two maxima (minima) per basis are denoted as $\max_1$ and $\max_2$ ($\min_1$ and $\min_2$) for calculating the visibility vis in the $\sigma_x$ and $\sigma_y$ bases. <u>A:</u> Raw coincidences for $\ell=1$. Coincidences for each angular position at the receiver are measured for 20 seconds. <u>B:</u> For $\ell=2$, we subtract accidental counts. Here, the coincidence counts for each angular position are measured for 40 seconds.

At the receiver on the rooftop of our institute IQOQI Vienna, we use a Newton-type telescope with a primary mirror of 37 cm diameter and a focal length of f=1,2m. In front of the primary mirror, we use an absorptive mask with a transparent, symmetric slit pattern to measure the modes (Fig. 2). The technique [5] allows us to measure visibilities in OAM-superposition bases, which is sufficient to verify entanglement. The masks are 40 cm in diameter and have a slit opening angle of 16° and 5.6° (for $\ell=1$ and $\ell=2$, respectively). The transmitted light is then detected on an APD with an active area of 500μm diameter. Similarly to the detection of the polarized partner photons, the arrival times are time tagged with a second TTM. To synchronize the time stamps on the two remote locations in the sub-nanosecond regime, we directly use the time-correlation of the photon pairs (which is inherently below 1ps), as explained in [30].

In the experiment, we perform visibility measurements in two mutually unbiased bases (MUBs). For photon A, the bases are diagonal or anti-diagonal and right- or left-circular polarization OAM ($|\psi_x\rangle = |H\rangle \pm |V\rangle$ and $|\psi_y\rangle = |H\rangle \pm i|V\rangle$). For photon B, we measure in the superposition-bases of two opposite OAM (specifically, $|\psi_x\rangle = |+\ell\rangle \pm |-\ell\rangle$ and $|\psi_y\rangle = |+\ell\rangle \pm i|-\ell\rangle$). Here, the superposition structure is a ring with $2\ell$ intensity maxima and minima. By changing the phase of the superposition, the intensity structure is rotated. With the slit mask, we can measure photons in the $\sigma_x$ and $\sigma_y$ bases, which correspond to the $\sigma_x$ and $\sigma_y$ bases of photon A [5]. Specifically, we measure fringes in coincidence counts for changing angular positions of the mask. We find minima and maxima of coincidences (Fig.3), and calculate the visibility $\text{vis} = \frac{\max_1 - \min_1 + \max_2 - \min_2}{\max_1 + \min_1 + \max_2 + \min_2}$ (max and min are highlighted with circles in Fig.3). To verify entanglement, we use an entanglement witness which is the sum of the two visibilities in the two MUBs [31]

$$W = \text{vis}_x + \text{vis}_y \begin{cases} \leq 1: \text{separable} \\ > 1: \text{entangled} \end{cases} \qquad (2)$$

All separable quantum states can reach at most W=1, which can be understood intuitively: If a product state is perfectly correlated in one basis, it cannot be correlated in any other mutually unbiased bases. Any experimental value above W=1 verifies entanglement in the system (a maximally entangled quantum state can have perfect visibility in both bases, thus W=2). The visibilities are calculated directly from the maxima and minima of the measured coincidences (blue/red and yellow/green circles in Fig. 3).

In the first measurement, we use the first higher order mode with $\ell=1$. We accumulate coincidence detections over 20 seconds at 20 different angular positions of the mask with a resolution of 9° (Fig. 3). The coincidence window is 2.5ns. Without any corrections (such as accidental coincidence subtraction) and without any assumption about the photon statistics, we get

$$W_{\ell=1} = 1.3644 \pm 0.0084,$$

which statistically significantly confirms entanglement between the two distant photons. The error stands for the standard deviation of the mean. We calculate the error by dividing the 20 second interval at each measurement position into 20 sections of equal length, and calculate the witness eqn.(2) for each of the 20 sections individually. From the resulting 20 values for $W_{\ell=1}$ (see [43]), we calculate the mean value and its uncertainty. In order to calculate the uncertainty of $W_{\ell=1}$, we did not need to assume any specific photon statistics. In many cases, Poissionian distribution is a good approximation of the photon statistics. However, it neglects additional sources of fluctuations, which can become relevant in experiments without controlled environments, such as free-space experiments. If we had assumed Poissionian distribution in our experiment, we would have underestimated the uncertainty significantly by around 70%, which is mainly due to atmospheric turbulences and instabilities at the sender. If we subtract accidental counts (~85±3 counts/sec), the average visibility in both bases will be roughly 84.2%. The visibility is enough to violate a Bell-type inequality, which would lead to violation of local realism and to the possibility of device-independent quantum key distribution.

In a second experiment, we transfer the photon to $\ell=2$ before transmission, send it to the receiver and measure coincidence counts for 20 different mask positions, each 4.5° rotated, for 40 seconds per setting. Here we get

$$W_{\ell=2} = 1.139 \pm 0.021,$$

verifying entanglement with $\ell=2$. Again, the error stands for the standard deviation of the mean, which has been evaluated equivalently as before: 40 seconds measurement intervals are divided into 40 parts of 1 second length. It results in 40 independent values of $W_{\ell=2}$ (see [43]) from which the mean and its error was obtained. Note that again we did not assume any information about the photon statistics. However, we had to subtract accidental coincidence counts (~95±7/sec, but measured for each measurement setting individually, see [43]), because the signal-to-noise ratio was significantly higher (5,5% for $\ell=2$ compared to 36,0% for $\ell=1$), which can be understood from the different weather conditions during that night, different detection efficiency at the telescope and smaller slit size of the masks (more details in [43]). As we collect timing information of the photons at the sending and receiving stations, we can access the number of accidental counts directly: At the offset between the time-tagging clocks, which correspond to the arrival times of the photons from a pair at the two locations, real coincidence counts from the entangled pairs are found. At every other offset, accidental coincidences can be seen (see [43]).

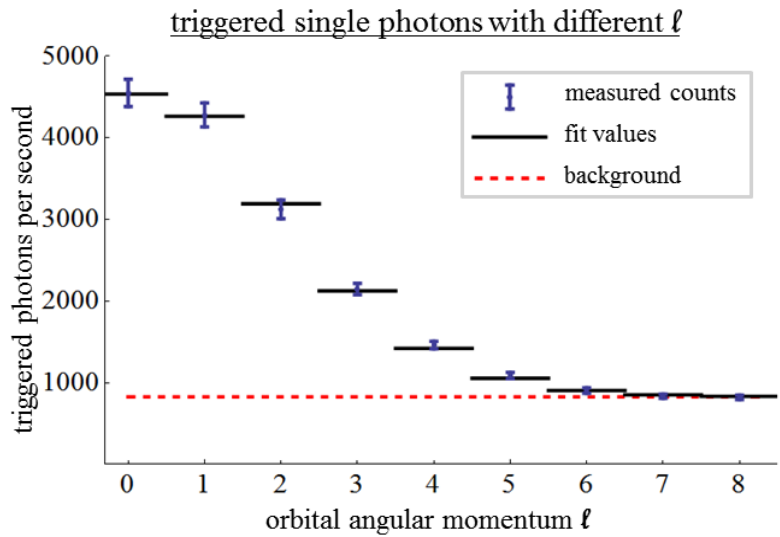

**Figure 4:** Triggered single photon counts for different orbital angular momentum $\ell$. We use correlated photon pairs (blue, each point is measured for 60 seconds) to determine the number of accessible OAM modes at the telescope. For that, we measure one photon at the sender, and transfer the correlated partner photon to a higher-order OAM mode (from $\ell$=0 to $\ell$=8), which is then transmitted to the telescope 3 kilometers away. The lower rate of coincidence counts for higher-order modes is due to the geometric restrictions (finite size of primary and secondary mirror) of the telescope, which can be modeled very well (black line [43]). The error bars show the standard deviation. The red dashed line indicates the background counts (calculated from average counts of $\ell$=10 to $\ell$=15). The data shows that we are able to access modes up to $\ell$ = 5 from the background, which constitutes 11 orthogonal quantum channels ($\ell$=-5 to $\ell$=5).

In both experiments for $\ell$=1 and $\ell$=2, we find visibilities smaller than unity. The reasons are an imperfect entanglement source (due to lack of temperature and vibration stabilization), imperfect detection method (the mask method can only give unity visibility for infinitesimal small slits), imperfect polarization compensation in fibers, accidental coincidence counts and atmospheric influence. The first order atmospheric influences are tip and tilt of the beam, which leads to relative misalignment between the mode and the mask. As the detection method is axis-dependent, it results in a significant drop in visibility (see [43]) which is larger for higher-order modes. However, that effect could be compensated with readily available adaptive optics.

Having confirmed that entanglement encoded in OAM can be transmitted over an intra-city link, we estimate the number of different orthogonal quantum channels we have access to in principle. As OAM modes grow for higher numbers of $\ell$ and our receiver telescope has a finite size, there is a maximum number of $\ell$ that can be detected. For that, we transfer photon B to different values of $\ell$ and (from $\ell$ =0 to 15) and record the number of coincidences with photon A. Thus, photon A is a trigger for the higher-order $\ell$-modes of photon B after sending it across the city. Here, no mask is in front of the telescope. The detected coincidence rates in Fig.4 show that photons up to $\ell$ =5 can be distinguished from the background. The graph can very well be described by the geometry of our telescope, which cuts the incoming beam both at the primary and secondary mirror [43]. We consider the counts of high order ($\ell \geq 10$) as background, as they reach an asymptotic value [43]. With our sender and receiver, we have access to roughly 11 quantum channels of OAM ($\ell$ =0 to $\ell$ =±5).

In conclusion, we are able to verify quantum entanglement of photon pairs with spatial modes over a turbulent, real-world link of 3 kilometers across Vienna. It shows that the spatial phase structure of single photons is preserved sufficiently well to be used in quantum optical experiments involving entanglement. By using the first two higher-order structures ($\ell=\pm1$, $\ell=\pm2$), we show that at least four additional orthogonal channels (in addition to the zeroth order Gaussian [1] case for $\ell=0$) permit long-distance quantum communication. While we still use two-dimensional subspaces, our result clearly shows that entanglement encoded in OAM can be identified after long-distance transmission. It is not fundamentally limited by atmospheric turbulences, as expected in some recent investigation, thus could be a feasible way to distribute high-dimensional entanglement.

We also show that our quantum link allows up to 11 orthogonal channels of OAM. The restrictive factors towards higher number of channels and higher quality of entanglement detection are technical limitations. The number of accessible channels can be increased by utilizing optimal generations of the modes (leading to smaller intensity structures) [32, 33] and larger sending and receiving telescopes. The quality of disturbed spatial modes can be improved with well-established adaptive phase-correction algorithms [34-36], which might lead to significantly larger quality in the entanglement identification. Adaptive measurement algorithms form another method to improve the entanglement detection, by adjusting the measurements according to the turbulence [37].

Entanglement of high-order spatially encoded modes over long distance opens up several interesting directions: Firstly, twisted photons have a large state space, and thus can carry more information than the well-studied case of polarization. Higher information capacity could be interesting for both classical and quantum communication, for example to increase the data rate. Additionally, in quantum key distribution [38-40] it could be used for increasing the robustness against noise, or improving the security against advanced eavesdroppers [41]. Secondly, OAM of photons permit complex types of entanglement due to their large state space. It also represents a physical quantity which can be (in principle) arbitrary large, thus it might be a very interesting testbed for fundamental tests. As such, curious phenomena such as the coupling of OAM modes with the space-time metric have been proposed [42]. We believe that our results will motivate both further theoretical and experimental research into the promising novel direction of long distance quantum experiments with twisted photons.

## Acknowledgement

We thank Roland Potzmann and ZAMG for providing access to the radar tower and detailed weather information. We also thank Mehul Malik for help with the experiment, Thomas Scheidl and Rupert Ursin for useful discussions and Nora Tischler for useful comments on the manuscript. This project was supported by the Austrian Academy of Sciences (ÖAW), the European Research Council (ERC Advanced Grant No. 227844 "QIT4QAD" and SIQS Grant No. 600645 EU-FP7-ICT), the Austrian Science Fund (FWF) with SFB F40 (FOQUS) and the Federal Ministry of Science, Research and Economy (BMWFW).

# Supplementary Information

## 1) Weather conditions during the measurement nights

Figure S1 depicts the detailed weather data from the measurement nights. Specifically important is Fig.S1A, the metrological visibility. It is a measure of the distance at which a black object can be clearly identified from the background and is defined as the distance $x_V$, where the contrast $C_V(x) = \frac{F_{BG}(x) - F_O(x)}{F_{BG}(x)} = 0{,}02$. $F_{BG}$ and $F_O$ are intensities of the background and the black object, respectively. At the position of the object (at x=0), $F_O(0)=0$, thus $C_V(0)=1$. Under very good conditions, $x_V$ can reach $\sim 100$km. In our experiment, the metrological visibility is a measure both of loss and background light. The loss can also be understood by large values of humidity, and background light is increased for low cloud base (Fig. S1F). The change in temperature (Fig. S1B) as well as the average and maximum wind (Fig. S1D,E) are influencing the turbulences of the atmosphere.

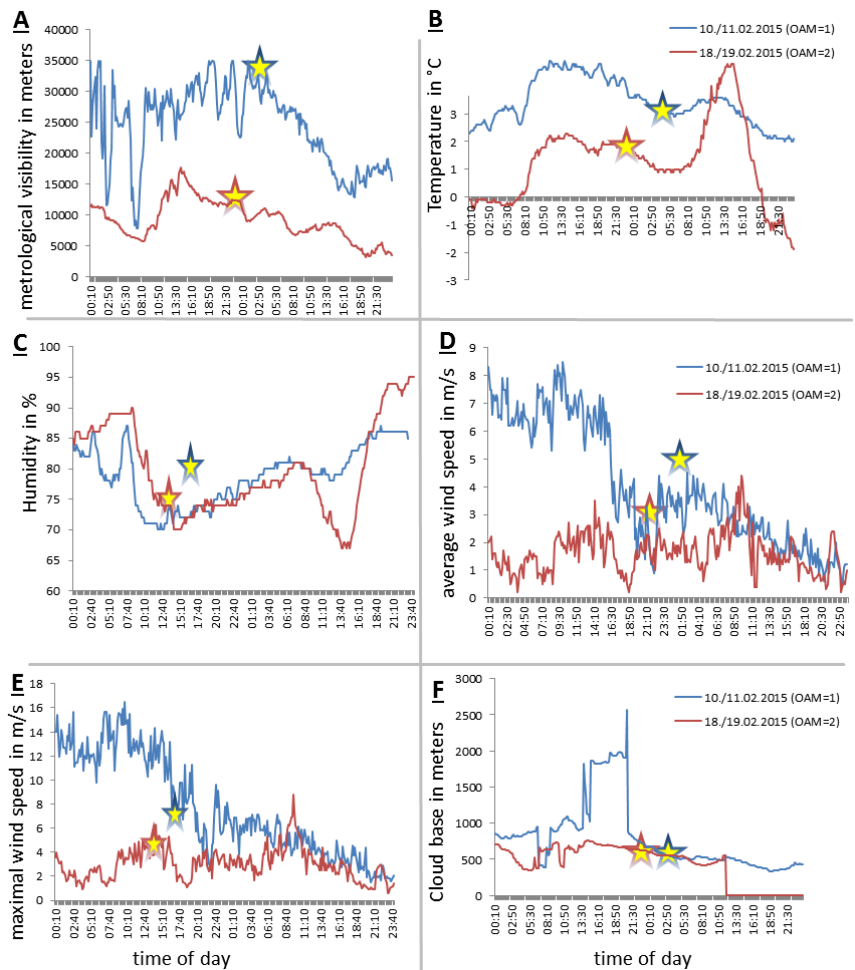

**Figure S1**: Detailed weather conditions of the two measurement nights. Blue (red) curve stands for the nights we measured entanglement involving OAM=1 (OAM=2). The blue (red) star indicated the precise time. A shows the metrological visibility, B shows the temperature, C shows the humidity, D shows the average wind speed, E shows the maximal wind speed and F shows the cloud base.

In the measurement of entanglement involving ℓ=2, the metrological visibility was significantly smaller which was one reason for the lower fringe visibility in Figure 2. Temperature, wind speed, humidity and cloud base were similar in the two nights.

## 2) Single photon counts and accidental coincidence counts

In the measurement of $\ell = 1$ and $\ell = 2$, we detect on average the following single photon counts:

|   | Sender (counts per Arm) | | | Receiver | | |
|---|---|---|---|---|---|---|
| $\ell$ | Singles/sec | Background/sec | Signal/Noise | Singles/sec | Background/sec | Signal/Noise |
| $\ell = \pm1$ | 800.000 | 10.000 | 79 | 36.600 | 27.800 | 0,36 |
| $\ell = \pm2$ | 770.000 | 10.000 | 76 | 40.100 | 38.000 | 0,055 |

**Table S1**: Single counts and Background counts at the sender (ZAMG) and receiver (IQOQI) for the experiment with $\ell = 1$ and $\ell = 2$.

In the measurement of entanglement with $\ell = 2$, we needed to subtract accidental coincidence counts. The reason was a large number of background counts compared to the small number of photons from the entangled pairs, with a signal-to-noise ratio of 0,055. The smaller ratio (compared to $\ell = 1$) can be understood by the smaller metrological visibility (as explained above), less efficient collection at the telescope (Fig. 4) and a smaller slit opening angle of the mask.

As we record the timing information of arriving photons (with time-tagging modules), we know the accidental count rates very accurately. The reason is that we implicitly measure the rate (for every real coincidence measurement) very often: Consider that we set the relative timing delay between the different time tagging clocks to zero, we see correlated photons. However, if we compare any other timing interval, we only see the accidental coincident counts. An example can be seen in Fig. S2. (The implicit assumption is that the accidental counts are independent of $\Delta t$.)

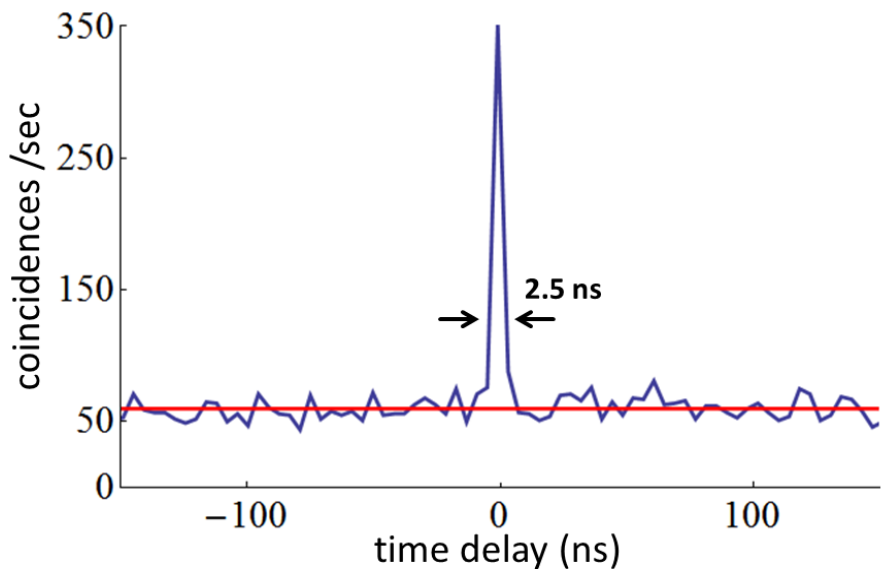

**Figure S2**: Example for measured coincidence counts from entangled photon pairs and accidental coincidence counts. The blue line indicates the coincidences calculated by overlapping the time-tagging files from the two locations. If the time delay is zero, real coincidences can be seen. For wrong time delays, only accidental counts are visible. The red line shows the average accidental counts.

## 3) Mode misalignment at mask

The detection method used here is axis-dependent, which means that the mode and the mask have to be aligned well (see Fig. 2B). Misalignments such as shifts of the mode relative to the mask will reduce the observable visibility. The misalignment can be introduced by first-order atmospheric influences such as tip and tilt. The effect can be simulated (Fig. S3), and we find that higher modes are more sensitive to alignment, which is one reason why the observed visibility for our $\ell = \pm 2$ measurement was lower than for the $\ell = \pm 1$ measurement.

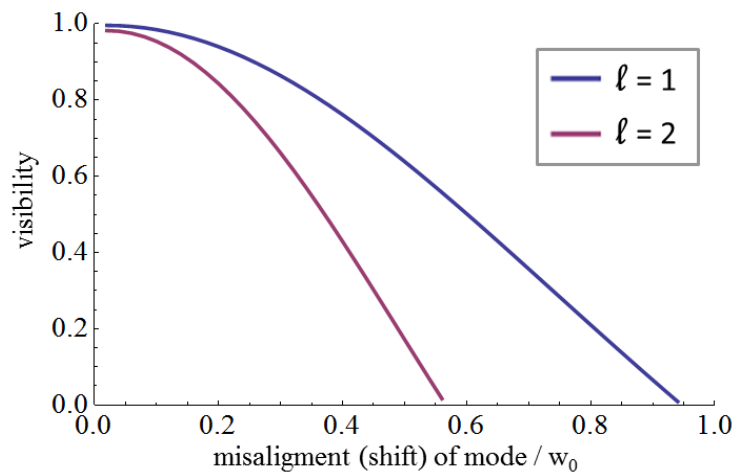

**Figure S3**: Visibilities of superpositions of $\ell = \pm 1$ and $\ell = \pm 2$ are plotted as a function of the relative misalignment of the mode and the mask, in units of the beam waist $w_0$. The higher-order mode is more sensitive to misalignment, which means the visibility drops faster. The visibilities are obtained by simulating a misaligned mode at a mask, and fitting the result with the expected sinusoidal function.

## 4) Witness values for 1 second intervals

The statistical error is calculated by separating the measured data into 1-second intervals, calculating the witness according to eqn.(2) and calculating the standard deviation of the mean.

For $\ell=1$, we have 20 witness values, because we measured each value for 20 seconds (sorted):

```
W=1.4354, 1.4214, 1.3942, 1.3940, 1.3928, 1.3858, 1.3848, 1.3841, 1.3804, 1.3692,
  1.3633, 1.3560, 1.3522, 1.3472, 1.3366, 1.3361, 1.3346, 1.3279, 1.3173, 1.2743.
```

For $\ell=2$, we have 40 witness values, because we measured each value for 40 seconds (sorted):

```
W=1.4218, 1.4201, 1.4106, 1.3708, 1.2581, 1.2570, 1.2452, 1.2338, 1.2245, 1.2180,
  1.2158, 1.2142, 1.2055, 1.2051, 1.2023, 1.2023, 1.1919, 1.1738, 1.1454, 1.1312,
  1.1235, 1.1091, 1.0995, 1.0950, 1.0778, 1.0702, 1.0695, 1.0524, 1.0492, 1.0410,
  1.0283, 1.0248, 1.0230, 1.0190, 1.0141, 0.9927, 0.9926, 0.9789, 0.8963, 0.8635.
```

## 5) Telescope geometry and OAM-dependent receiving of photons

The telescope cuts the beam in two different ways, which result in a reduced number of detected photons: Firstly, the primary mirror has a finite diameter of roughly 37 cm. As OAM modes increase in size if one increases the $\ell$ value, higher order modes are more significantly cut. Secondly, the secondary mirror of the telescope is in the centre of the beam path. The Gauss mode has the maximal intensity there, thus it is substantially cut at the secondary mirror. The effect can be illustrated graphically:

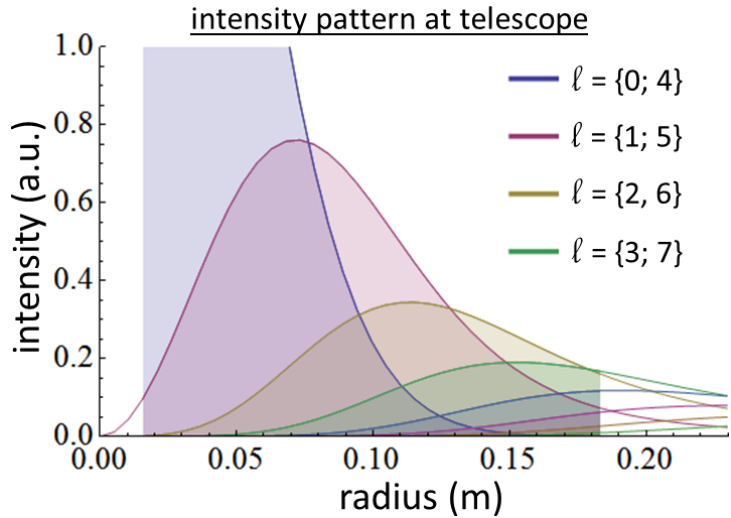

**Figure S4**: The blue, red, yellow and green curves present the radial intensity distribution of OAM modes $M_\ell(r,\varphi=0)^2$. ($\ell$={0;4} means that blue stands for $\ell$=0 and $\ell$=4). The filled region indicates the part of the beam that arrives at the photon detector. It is limited on the inside by the shading of the secondary mirror and on the outside by the finite size of the primary mirror.

In the experiment, we prepare OAM-modes with phase-only holograms. The size of those modes scale linear with $\ell$. There is no closed-form solutions for such holograms, thus we calculate them numerically by applying a Fourier transformation of a Gaussian beam with a helical phase (which is introduced by the SLM):

$$M_\ell(r,\varphi) = N \cdot \mathcal{F}\left[ e^{-\frac{r^2}{w_0^2}} \cdot e^{i\ell\varphi} \right] \tag{S1}$$

Here, $\mathcal{F}[.]$ stands for the Fourier-Transformation, $w_0$ is the beam-waist and N is a normalization constant. The expected intensities for different modes can be easily calculated with the given boundary conditions. The function

$$F(\ell) = \text{bg} + \text{cnt} \cdot \int_0^{2\pi} d\varphi \int_{SM}^{PM} r \cdot M_\ell(r,\varphi)^2 dr \tag{S2}$$

is used to obtain the fit in Fig. 4, with cnt (counts), SM (size of secondary mirror), and $w_0$ (beam waist at telescope) being the fit-paramters. Here, the background bg has been calculated from the measured data (the mean of the counts from $\ell$=10 to $\ell$=15), and PM (size of primary mirror) has been measured to be roughly 18,5 cm. The obtained fit value resemble very well the physical dimensions of the telescope (SM=1,2cm) and of the observed classical alignment beam ($w_0$=8,5cm) in Fig. 2.

## 6) Triggered single photons and background

In Fig. 4, we show the potential of our telescope to detect high-order modes, which is restricted due to the finite size of the primary and secondary mirror (see Fig. S3 above). In that experiment, we are not interested in entanglement, but only in the number of detected triggered single photons in higher-order modes. Therefore, we removed the polarizer after the transfer-setup to double the number of photons that are sent. We create the state

$$|\psi\rangle = \tfrac{1}{\sqrt{2}}\big(|H,0\rangle_A|V,0\rangle_B + |V,0\rangle_A|H,+\ell\rangle_B\big), \tag{S3}$$

where A stands for photon A, which is detected at the sender, and B stands for photon B which is sent across the city to the receiving telescope. The first term $|H,0\rangle_A|V,0\rangle_B$ is used to actively synchronize the time-tagging modules at the two locations (as explained in the main-text). The second term is what we are measuring for Fig.4: Whenever the photon at the sender is vertically polarized, we know that we transmitted photon B with $\ell$, where $\ell$ goes from 0 to +15.

The background in Fig.4 originates mainly from two effects: The first effect is accidental coincidence counts because of large count rates (1.500.000 single counts/sec at the receiver, which leaded to $\sim 525 \pm 10$ accidental counts/sec). The second effect is due to misidentification of polarization in the transfer-setup ($\sim 170 \pm 25$ counts/sec). If the polarization changes between the source and the transfer-setup (which is likely because fibers where used, which are not compensated perfectly), some photons end up in a Gaussian state, which are detected by the telescope. For large values of $\ell$ ($\ell \geq 10$) the counts reach an asymptotic value, which we then define as our background.